\documentclass[aps,prd, onecolumn,groupedaddress,nofootinbib,preprintnumbers,superscriptaddress]{revtex4-2}  

\usepackage{graphicx,mathtools,tabu,array}
\usepackage{epstopdf}
\usepackage{amsmath}
\usepackage{amsfonts}
\usepackage[colorlinks=true,citecolor=red,linkcolor=blue,breaklinks=true]{hyperref}
\usepackage{accents}
\usepackage[figure, table]{hypcap}
\usepackage{amssymb}
\usepackage{appendix}
\usepackage{comment}
\usepackage{bbold}
\usepackage{xcolor}
\usepackage{slashed}
\usepackage{subfigure}
\usepackage{setspace}
\usepackage{footnote}
\usepackage{multirow}
\usepackage{braket}
\usepackage[normalem]{ulem}
\usepackage[utf8]{inputenc}
\usepackage{mathrsfs}
\usepackage{longtable}
\usepackage{enumitem}
\usepackage{booktabs} 
\usepackage{subfigure}

\begin{document}

\preprint{IFT-UAM/CSIC-25-17}

\title{Gray-body factors: Method matters}

\author{Alexandre Arbey}
\email{alexandre.arbey@ens-lyon.fr}
\affiliation{Université Claude Bernard Lyon 1, CNRS/IN2P3, Institut de Physique des 2 Infinis de Lyon, UMR 5822, F-69622, Villeurbanne, France}

\author{Marco Calz\`a}
\email{marco.calza@unitn.it}
\affiliation{Department of Physics, University of Trento, Via Sommarive 14, 38123 Povo (TN), Italy}
\affiliation{Trento Institute for Fundamental Physics and Applications (TIFPA)-INFN. Via Sommarive 14, 38123 Povo (TN), Italy}

\author{Yuber F. Perez-Gonzalez}
\email{yuber.perez@uam.es}
\affiliation{Departamento de F\'{i}sica Te\'{o}rica and Instituto de F\'{i}sica Te\'{o}rica (IFT) UAM/CSIC, Universidad Aut\'{o}noma de Madrid, Cantoblanco, 28049 Madrid, Spain}

\begin{abstract}

    The calculation of gray-body factors is essential for understanding Hawking radiation and black hole thermodynamics.  
    While the formalism developed by Chandrasekhar is effective for static black holes, it faces significant challenges in Kerr spacetimes, particularly in the superradiant regime, where a specific choice of coordinates introduces numerical inaccuracies. 
    To address these limitations, an alternative method based on re-scaling radial coordinates and employing Frobenius-like expansions has been investigated.  
    We compare the gray-body factors obtained for a near-maximally rotating black hole using both methods and find that the Chandrasekhar formalism systematically overestimates the values in the superradiant regime compared to well-established analytical results.  
    Specifically, for a spin parameter of $a_* = 0.999$, the Chandrasekhar method yields values approximately twice as large as the correct result.  
    Since this approach has been implemented in \texttt{BlackHawk}, we assess the impact of these discrepancies on constraints derived from gamma-ray observations of highly spinning primordial black holes.
\end{abstract}

\maketitle

\section{Introduction}

\noindent Black holes, as some of the most enigmatic objects in the universe, offer a unique window into fundamental physics. They bridge the realms of classical gravity, quantum field theory, and thermodynamics, serving as natural laboratories for exploring the interplay between these domains. Among the various types of black holes, Kerr black holes, characterized by their rotation, play a particularly significant role. Their rich phenomenology, including the superradiant amplification of scattered waves and their thermodynamic properties, has captured the attention of physicists for decades. Furthermore, the study of black holes has transcended theoretical interest, finding applications in astrophysics, cosmology, and even potential observational signatures.

\noindent A particularly intriguing application arises in the context of Hawking radiation, a semi-classical phenomenon through which black holes emit particles, leading to a gradual loss of mass and eventual evaporation~\cite{Hawking:1974rv,Hawking:1974sw}. This effect not only connects black holes to quantum field theory but also makes them potential sources of observational signals in the form of high-energy photons or other particles. The emitted spectrum, however, deviates from a perfect black-body curve due to the presence of Gray-Body Factors (GBFs), which encode the scattering and absorption properties of the black hole. These GBFs depend on the geometry, spin, and other characteristics of the black hole, making their accurate computation a critical task for understanding black hole emissions. Although GBFs historically found their main application in Hawking radiation~\cite{Hawking:1974rv,Hawking:1974sw} and in the absorption cross section, recently their connection with ring-down spectral amplitudes~\cite{Rosato:2025byu,Oshita:2024fzf,Rosato:2024arw,Oshita:2023cjz,Oshita:2022pkc} and Quasi-Normal Modes (QNMs)~\cite{Konoplya:2024lir,Konoplya:2024vuj} was shown.

\noindent 
In recent years, primordial black holes (PBHs) have gained significant attention due to their potential impact on cosmic evolution and the possibility that they constitute dark matter (DM) \cite{Green:2020jor,Carr:2020gox,Arbey:2021gdg,Auffinger:2022khh,Arbey:2024ujg}. The detection of gravitational waves (GWs) and astrophysical black holes~\cite{LIGOScientific:2016aoc} has further fueled interest in their existence. Strong constraints on PBH abundance exist across different mass ranges~\cite{Carr:2020gox,Carr:2020xqk}, with those below a few times $10^{14}~{\rm g}$ having either fully evaporated or nearing their final stages today, assuming only Standard Model (SM) degrees of freedom.

\noindent PBHs have been widely studied as probes of new physics \cite{Federico:2024fyt,Maggio:2022nme,Maggio:2021ans,Calza:2024ncn,Calza:2023iqa,Calza:2022ioe,Calza:2021czr,Calza:2023rjt,Horowitz:2023xyl,Vagnozzi:2022moj,Afrin:2022ztr,Baker:2021btk,Baker:2022rkn}. Considering Hawking evaporation as source of particles, PBH could have emitted the observed DM. In fact, even if only interacting gravitationally~\cite{Lennon:2017tqq,Morrison:2018xla,Hooper:2019gtx,Auffinger:2020afu,Gondolo:2020uqv,Bernal:2020bjf,Bernal:2020ili,Bernal:2020kse,Baldes:2020nuv,Masina:2020xhk,Masina:2021zpu,Sandick:2021gew,Bernal:2021bbv,Bernal:2021yyb,Cheek:2021odj,Cheek:2021cfe,Barman:2021ost,Bernal:2022oha,Cheek:2022mmy,Chen:2023lnj,Chen:2023tzd,Kim:2023ixo,Haque:2023awl,RiajulHaque:2023cqe,Chaudhuri:2023aiv,Gehrman:2023esa,Gehrman:2023qjn,Carr:2016drx,Clesse:2016vqa,Carr:2017jsz,Green:2020jor,Croon:2020ouk},  
they could contribute to Dark Radiation~\cite{Hooper:2019gtx,Lunardini:2019zob,Masina:2020xhk,Masina:2021zpu,Hooper:2020evu,Arbey:2021ysg,Cheek:2022dbx,Papanikolaou:2023oxq},  
and potential drivers of baryogenesis~\cite{Barrow:1990he,Majumdar:1995yr,Upadhyay:1999vk,Dolgov:2000ht,Bugaev:2001xr,Baumann:2007yr,Hamada:2016jnq,Hooper:2020otu,Gehrman:2022imk}  
or leptogenesis~\cite{Fujita:2014hha,Hooper:2020otu,Perez-Gonzalez:2020vnz,Bernal:2022pue,JyotiDas:2021shi,Calabrese:2023key,Calabrese:2023bxz,Schmitz:2023pfy,Ghoshal:2023fno,Datta:2020bht,Gunn:2024xaq,Calabrese:2025sfh}.  
They have also been associated with gravitational wave production~\cite{Papanikolaou:2020qtd,Domenech:2020ssp,Bhaumik:2022pil,Bhaumik:2022zdd,Papanikolaou:2022chm,Ghoshal:2023sfa}  
and the stability of the SM Higgs potential~\cite{Burda:2016mou,Burda:2015isa,Hamaide:2023ayu}.  
Given these wide-ranging implications, accurate modeling of PBH emission spectra, including the effects of GBFs, remains essential for constraining their abundance and understanding their role in cosmic evolution.

\noindent The computation of GBFs, however, is fraught with challenges, especially in the case of rotating black holes. For non-rotating, spherically symmetric black holes, the Schr\"odinger-like formalism simplifies the problem, reducing it to solving a one-dimensional scattering equation with known asymptotic solutions. If spherical symmetry is preserved and no superradiant regime is present this method allows the solution of the problem for a wide number of BHs \cite{Calza:2025whq,Arbey:2021jif,Arbey:2021yke}. This approach, while effective, encounters significant difficulties when extended to Kerr black holes, where the rotational dynamics introduce complexities such as the coupling of modes and the phenomenon of superradiance. Moreover, the use of specific tortoise coordinates (introduced by Chandrasekhar and Detweiler \cite{Chandrasekhar:1975zx,Chandrasekhar:1976zz,Chandrasekhar:1977kf}, and for now on dubbed as Chandrasekhar tortoise coordinates), a key component of the Schr\"odinger-like formalism, introduces numerical singularities and inaccuracies that can propagate into the resulting GBFs.

\noindent To address these issues, an alternative formalism is considered, aiming to circumvent the challenges posed by the Chandrasekhar tortoise coordinate and to provide more accurate descriptions of wave propagation in the Kerr geometry. These methods, which leverage re-scaled radial coordinates and advanced expansions, offer a promising avenue for improving the accuracy of GBF calculations, particularly for high-spin black holes where traditional approaches falter. Beyond their theoretical importance, these refinements have direct implications for observational studies, as they influence predictions of high-energy particle spectra and their potential detectability.

\noindent This work revisits the computation of GBFs for Kerr black holes, comparing the Schr\"odinger-like formalism with an approach that avoid the use of Chandrasekhar tortoise coordinates. Both methods are present in the literature and extensively used since the 1970's. By analyzing their respective accuracies and applicability, we aim to identify the most robust methods for studying black hole radiation, with a particular focus on the high-spin regime. These efforts contribute to a more comprehensive understanding of black hole physics and their potential observational signatures, with implications spanning astrophysics, cosmology, and beyond.

\noindent This work is organized as follows.
In Sec.~\ref{sec:hawk_sp}, we briefly review the Hawking spectrum, emphasizing the role of GBFs.  
Secs.~\ref{sec:sch_form} and \ref{sec:bet_form} introduce the two main approaches considered in this work.  
Sec.~\ref{sec:sch_form} examines the Schr\"odinger-like formalism, where the radial component of a field in the Kerr metric —- described by the Teukolsky equation —- is transformed into a Schr\"odinger-like differential equation.  
Sec.~\ref{sec:bet_form} presents an alternative approach based on the direct solution of the Teukolsky equation.  
Our results are presented in Sec.~\ref{sec:results}, where we compare GBFs, the Hawking spectrum, and PBH constraints on their contribution to DM, obtained using both methods.  
Throughout this work, we adopt Planck natural units, setting $\hbar = c = k_{\rm B} = G = 1$.

\section{Hawking Radiation Spectrum}\label{sec:hawk_sp}

\noindent As first demonstrated in Refs.~\cite{Hawking:1974rv,Hawking:1974sw}, a black hole emits particles of type $i$ with spin $s$ and $g_i$ internal degrees of freedom, following a thermal spectrum given by  
\begin{equation}\label{eq:PHR}
    \frac{d^2N_i}{dt d\omega_i}=\frac{g_i}{2\pi}\sum_{l,m}\frac{\,_s\Gamma^m_{l}(\omega)}{ e^{2\pi k(\omega)/\kappa}\pm 1}~,
\end{equation}
where $k = k(\omega)$ depends on the particle energy $\omega$ and additional black hole properties, and $\kappa$ is the black hole’s surface gravity. The plus/minus sign corresponds to the fermionic/bosonic statistic respectively.
In Eq.~\eqref{eq:PHR}, the coefficients $\,_s\Gamma^m_{l}$ represent the GBFs, which quantify deviations from a perfect black-body spectrum for each $(l,m)$ mode in a spheroidal wave decomposition.

\noindent To compute the gray-body factors $\,_s\Gamma^m_{l}$, one must solve the equation of motion for a given spin type in the black hole spacetime and determine the transmission coefficients by imposing ingoing boundary conditions at the event horizon.  
For spherically symmetric metrics, several well-established techniques exist for calculating these factors, see, e.g., Refs.~\cite{Arbey:2021jif,Arbey:2021yke,Calza:2025whq}. However, for more general black hole spacetimes, numerical challenges arise due to the loss of certain symmetries.  
In the following, we focus on the case of a black hole with a non-negligible angular momentum, where such difficulties become particularly relevant.

\section{Schr\"odinger-like formalism}\label{sec:sch_form}

\noindent The Newman-Penrose formalism~\cite{Newman:1961qr} provides a unified framework for expressing the field equations of different spin $s$ in a single equation, known as the Teukolsky equation~\cite{Teukolsky:1973ha,Press:1973zz,Teukolsky:1974yv}. 
After performing a variable transformation, the radial component of the Teukolsky equation can be rewritten in a particularly compact and convenient form~\cite{Chandrasekhar:1975zx,Chandrasekhar:1976zz,Chandrasekhar:1976bho,Chandrasekhar:1976ap,Chandrasekhar:1977kf,Chandrasekhar:1985}, commonly referred to as the Schr\"odinger-like form,
\begin{equation}\label{eq:sch_lk}
\partial_{r^*}^2\psi_s(r^*) + \left[\omega^2 - V_s(r(r^*))\right]\psi_s(r^*)=0,
\end{equation}
with $r^*$ the Chandrasekhar tortoise coordinate, which will be defined later on. 
The wave equations in this form offers a great simplification in the computation of the scattering problem since, usually, the potential $V_s(r^*)$ vanishes for $r^* \rightarrow \pm \infty$, and asymptotic solutions are known. 
Therefore, imposing purely in-going boundary conditions at the event horizon and integrating out, it is possible to compute the transmission coefficient or gray-body factors (GBFs).

For example, in the case of spherically  symmetric metrics
\begin{equation}
    ds^2=-F(r) dt^2 + \frac{dr^2}{G(r)} + H(r)d\Omega^2,
\end{equation}
with $d\Omega^2= d\theta^2 +\sin \theta d\phi^2$ and $F, G, H$ such that the metric is asymptotically flat, the choice 
\begin{equation}
    \frac{d r^*}{dr} = \frac{1}{\sqrt{ F G}}
\end{equation}
ensures that it is possible to recover Eq.\eqref{eq:sch_lk} with a vanishing potential at infinities \cite{Arbey:2021yke}.

\noindent Nonetheless, in the case of a rotating BH some complications emerge resulting in a less straightforward computation of the scattering problem.
Let us first introduce the Kerr metric, which in the Boyer-Lindquist coordinates $(t,r, \theta,\varphi)$ reads~\cite{Boyer:1966qh}
\begin{align} \label{eq:Kerr_metric}
    ds^2 = &- \left( 1- \frac{2 M r}{\Sigma}\right)dt^2 - \frac{4 M a r \sin^2 \theta }{\Sigma}dt d\varphi 
    +\frac{\Sigma}{\Delta}dr^2 + \Sigma d\theta^2 + \left(r^2 + a^2 + \frac{2 M a^2 r \sin^2 \theta}{\Sigma}\right) \sin^2 \theta d\varphi^2,
\end{align}
where $M$ is the hole mass, $a=J/M$ is the BH angular momentum, $\Delta=r^2 + a^2 - 2 M r$, and $\Sigma=r^2 a^2 \cos^2 \theta$.
This solution has an inner Cauchy horizon $r_-$ and an outer event horizon $r_+$, both roots of $\Delta$.

\noindent The Teukolsky equation for the radial component of a field with spin $s$, $R_s$, in the Kerr spacetime is given by
\begin{equation}\label{eq:radTeq}
\Delta^{-s} \partial_r(  \Delta^{s+1}\partial_rR_s) + [(K^2 -2is(r-M)K)\Delta^{-1}+4 i s \omega r -\, _s Q^m_l]R_s=0~,
\end{equation}
where $_s Q^m_l={_s A^m_l }+ a^2 \omega^2 - 2 a \omega m$ and $K=(r^2 + a^2) \omega - m a  $. ${_s A^m_l } = {_s A^m_l }(a\omega)$ are the eigenvalues of the spin-weighted spheroidal harmonics and cannot be expressed analytically in terms of the spherical angular momentum quantum numbers $l, m$. Nevertheless, for $a \omega \ll 1$ they can be computed using a perturbative expansion
 \cite{Berti:2005gp}, yielding:
\begin{eqnarray}\label{separcon}
{}_s A^m_l (a\omega) &=&l(l+1) - s(s+1) -a\omega \frac{2ms^2}{l(l+1)}\nonumber\\
&+& (a\omega)^2 \left\{\frac{2}{3} \left[ 1+ \frac{3m^2-l(l+1)}{(2l-1)(2l+3)}\right] - \frac{2s^2}{l(l+1)} \frac{3m^2-l(l+1)}{(2l-1)(2l+3)} \right.\nonumber\\
&+& \left.2s^2 \left[ \frac{(l^2-s^2)(l^2-m^2)}{l^3(2l-1)(2l+1)} - \frac{[(l+1)^2-m^2][(l+1)^2 - s^2]}{(l+1)^3(2l+1)(2l+3)}\right]\right\} \nonumber\\
&+&\mathcal{O}\left[(a \omega)^3\right]~.
\end{eqnarray}

\noindent As established in the Chandrasekhar and Detweiler series of papers~\cite{Chandrasekhar:1975zx,Chandrasekhar:1976zz,Chandrasekhar:1977kf}, to recover the form of Eq.~\eqref{eq:sch_lk} in the case of a Kerr background, one defines the $r^*$ coordinate as
\begin{equation}\label{eq:tort}
    \frac{dr^*}{dr}=\frac{\rho^2}{\Delta}
\end{equation}
with $\rho(r)^2=r^2 +a^2 - a m / \omega$, 
$m$ being the azimuthal number of the specific mode.
Integrating Eq.~\eqref{eq:tort} with a vanishing integration constant one has
\begin{equation}\label{eq:Ktort}
    r^*(r)=r+\frac{r_S r_+ - a m /\omega}{r_+ - r_-} \log\left(\frac{r}{r_+}-1\right)-\frac{r_S r_- - a m /\omega}{r_+ - r_-} \log\left(\frac{r}{r_-}-1\right),
\end{equation}
where $r_S=2M$ is the Schwarzschild radius, and $r_\pm$ are the event and inner BH horizon and read
\begin{equation}
    r_\pm=\frac{r_S}{2}\left(1 \pm \sqrt{1 - a_*^2}\right),
\end{equation}
with the dimensionless $a_\star$ is defined as $a_\star = a/M = J/M^2$.
Typically, the coordinate transformation $r^*(r)$ is an invertible function, varying monotonically from negative infinity at $r = r_+$ to positive infinity for large values of $r$, where $r \sim r^*$.  
When this happens the inverse function $r(r^*)$ exists over the entire domain $-\infty < r^* < +\infty$, allowing one to express the potential as $V_s(r(r^*))$ and proceed with the scattering problem. This approach was successfully applied in the previously studied spherically symmetric case.

\begin{figure}
    \centering
    \includegraphics[width=0.5\linewidth]{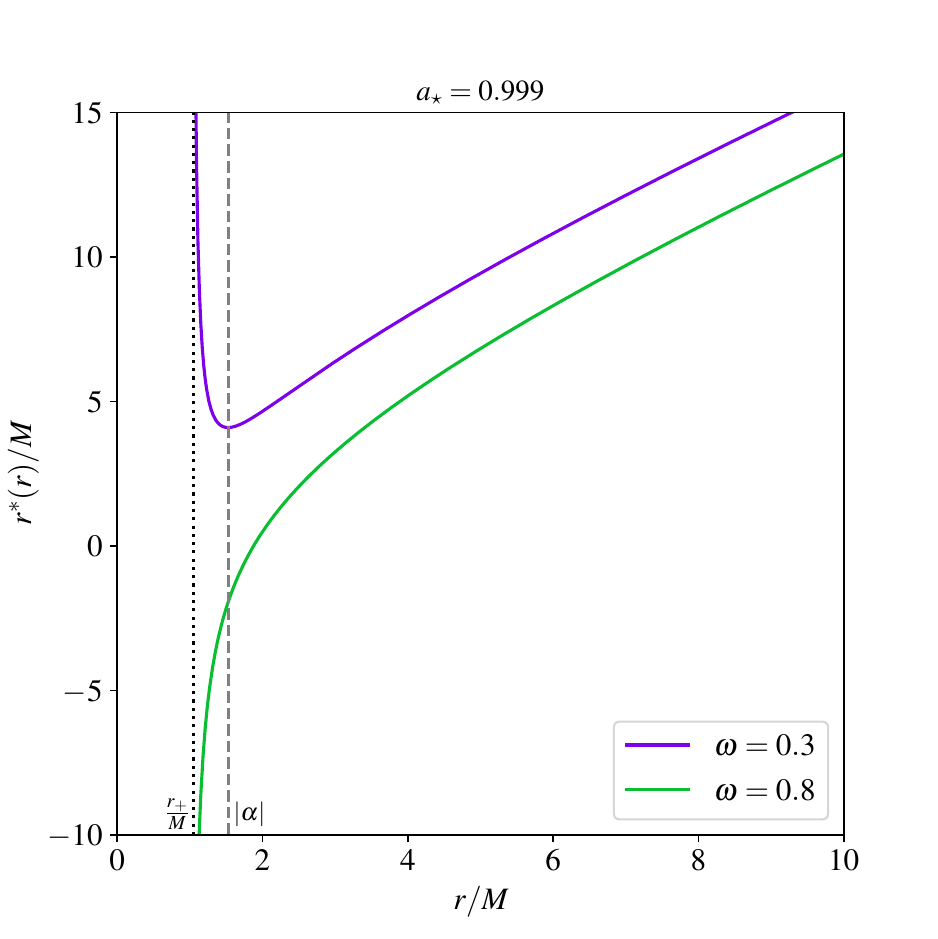}
    \caption{$r^*(r)$ in units of the black hole mass $M$ for a co-rotating mode with $m=1$ and energies $\omega=0.3$ (purple) and $\omega=0.8$ (green).
    The dotted line indicates the value of $r_+$, while the dashed corresponds the value of the minimum $|\alpha|$, where $\alpha^2 = a^2 - am/\omega$.}
    \label{fig:fig1}
\end{figure}

\noindent In contrast, applying this formalism to the Kerr spacetime introduces additional challenges.  
Specifically, in the superradiant regime, which occurs when $\omega < m \Omega_H$, where  
\begin{align}
    \Omega_H = \frac{a_\star}{r_S(1+\sqrt{1-a_\star^2})}
\end{align}
denotes the angular velocity of the black hole horizon, the gray-body factors become negative due to wave amplification effects that may arise for massive bosons.  
Moreover, in this regime, Eq.~\eqref{eq:Ktort} is not invertible for co-rotating modes (positive $m$) when the frequency satisfies $\omega < \omega_s$, with $\omega_s = am/(r_S r_+)$.  
This behavior is illustrated in Fig.~\ref{fig:fig1}, where we show the dependence of $r^*(r)$ for a co-rotating mode with $m=1$ for two cases: $\omega = 0.3$ and $\omega = 0.8$.  
As observed, for the lower energy case, $r^*(r)$ exhibits a minimum around $r=1.5 M$.  
This implies that $r^*(r)$ is not monotonically increasing, preventing the inversion of the relation needed for numerically solving the Teukolsky equation in its Schr\"odinger-like form.  
The effective potentials $V_s(r^*)$ for scalar, fermion, vector, spin-$3/2$, and spin-2 fields are given by~\cite{Chandrasekhar:1975zx,Chandrasekhar:1975zz,Chandrasekhar:1976bho,Chandrasekhar:1976zz,TorresDelCastillo:1989hk,TorresdelCastillo:1990aw,TorresdelCastillo:1992zq,Arbey:2019mbc,Arbey:2021mbl}:
\begin{subequations}
    \begin{align}
        &V_0(r) = \frac{\Delta}{\rho^2}\left(\,_0Q^m_l + \frac{\Delta+2r(r-M)}{\rho^2} - \frac{3r^2\Delta}{\rho^4}\right),\\
        &V_{1/2,\pm} = (\,_{1/2}Q^m_l + 1)\frac{\Delta}{\rho^4}\mp \frac{\sqrt{\,_{1/2}Q^m_l + 1}}{\rho^4}\left(r-M-\frac{2r\Delta}{\rho^2}\right),\\
        &V_{1,\pm} = \frac{\Delta}{\rho^4}\left[\,_{1}Q^m_l + 2 - \alpha^2\frac{\Delta}{\rho^4}\pm i\alpha \rho^2\frac{d}{dr}\left(\frac{\Delta}{\rho^4}\right)\right],\\
        &V_{3/2} = \frac{\Delta}{\rho^6}\left(q_{3/2}-\frac{ f_{3/2}(r)}{(q_{3/2}-\beta_{3/2}\sqrt{\Delta})^2}\right),\\
        &V_{2} = \frac{\Delta}{\rho^8}\left(q_2 - \frac{\rho^2 f_{2}(r)}{(q_2-\beta_2\Delta)^2}\right),
    \end{align}
\end{subequations}
where $q_{3/2} = \,_{3/2}Q^m_l \rho^2 + 2MR -2a^2$, $q_2 =(\,_{2}Q^m_l+4)\rho^4 +3\rho^2(r^2-a^2)-3r^2\Delta$, $\beta_{3/2}=\pm2\sqrt{a^2 - \,_{3/2}Q^m_l\alpha^2}$, and $\beta_2 = \pm 3\alpha^2$.  
The functions $f_{2}(r)$ and $f_{3/2}(r)$ are lengthy expressions provided in Refs.~\cite{Arbey:2019mbc,Arbey:2021mbl}, but they are not relevant for our discussion.  
These potentials exhibit singularities at $r^2=-\alpha^2$, where $\alpha^2 = a^2 - am/\omega$~\cite{Chandrasekhar:1975zx,Chandrasekhar:1977kf}. 
Additionally, for spin-$3/2$ and spin-2 fields, extra divergences appear for certain sign combinations. 

\noindent To address this issue, one needs to introduce a technique that involves inverting the function $r^*(r)$ in two branches, effectively bypassing the minimum at $r_{\rm min}=|\alpha|$. 
Specifically, one must impose purely ingoing boundary conditions at the horizon for $r^* \sim +\infty$ and integrate the solution along the first branch up to $|\alpha| - \epsilon$, with $\epsilon \rightarrow 0$. 
This intermediate solution is then used as a boundary condition for a secondary integration along the second branch, from $|\alpha|+ \epsilon$ to $+\infty$.
However, when the energy exceeds $\omega_s$, the relation becomes single-valued, allowing inversion and numerical computation.
This approach enables the treatment of Kerr black holes while preserving the wave equation in the form of Eq.~\eqref{eq:sch_lk}, significantly simplifying the mathematical formulation. Due to its effectiveness, this method has been incorporated into various publicly available codes~\cite{Arbey:2019mbc,Arbey:2021mbl, Perez-Gonzalez:2023uoi}.

\section{Improved Formalism for Rotating Black Holes}\label{sec:bet_form}

\noindent In this section, we analyze an alternative approach to determine the GBFs, which has been already discussed in the literature. 
However, to the best of our knowledge, it has not yet been recognized that this method provides a more accurate approximation for rotating black holes. 
By avoiding the use of the Chandrasekhar tortoise coordinates, it circumvents the spurious singularity introduced by them.
We can rewrite Eq.~(\ref{eq:radTeq}) in terms of the re-scaled radial coordinate $x=(r-r_+)/r_+$:
\begin{equation}\label{eq:Teuk2}
    x^2(x+\tau)^2 \partial_x^2 R(x) +(s+1)(2 x+\tau)x(x+\tau) \partial_x R(x) + V(x)R(x)=0\, ,
\end{equation}
where the effective radial potential $V(x)$ is 
\begin{equation}\label{eq:Near}
    V(x)=k^2- i s (2 x + \tau) k + (4 i s \omega (x+1)-\,_sQ^{m}_l)x (x+\tau)~,
\end{equation}
being $k=(2-\tau)(\omega-m \Omega_H) r_+ + x(x+2)\omega r_+$, with  $\tau=(r_+ - r_-)/r_+$. 

\noindent The imposition of purely ingoing boundary conditions near the horizon can be expressed as a Taylor expansion~\cite{Wilson1928AGS,Baber_Hassé_1935,Leaver:1985ax,Leaver:1986vnb,Konoplya:2023ahd,Konoplya:2023ppx,Rosa:2012uz,Rosa:2016bli}
\begin{eqnarray}\label{eq:frob}
R_s(x) = x^{-s - \frac{i \varpi}{\tau}} \sum_{n=0}^\infty a_n x^n\,,
\end{eqnarray}
which corresponds to a Frobenius series, a well-known method for solving second-order differential equations~\cite{Calza:2024fzo,Calza:2024xdh}. Here $\varpi=(2-\tau)(\bar \omega - m \bar \Omega_H)$, and with barred quantities being normalized to the horizon radius, e.g. $\bar \omega= \omega r_+$.
As described in Sec.~V of \cite{Teukolsky:1973ha} with the aid of the useful coordinates $\frac{d \tilde r}{dr}=\frac{r^2 + a^2}{\Delta}$ and some further manipulations, one obtains the asymptotic solutions of Eq.~\eqref{eq:radTeq}. Namely, $R_s$ in the vicinity of the event horizon reads 
\begin{align}
    R_s \sim \,_sR^{lm}_{\text{hole}}\, \Delta^{-s}\, e^{-i \bar \omega \tilde r}, \quad \quad (r \rightarrow r_+)\,,
    \label{eq:asymptotic}
\end{align}
While, in the asymptotic region far from the black hole horizon, it take the form  
\begin{align}
    R_s \sim \,_sR^{lm}_{\text{in}} \frac{e^{-i\omega r}}{r} + \,_sR^{lm}_{\text{out}} \frac{e^{i\omega r}}{r^{2s+1}}, \quad r\gg r_+\,.
\end{align}
Expressing the later in terms of the rescaled coordinate of Eq.~\eqref{eq:Teuk2}, we obtain  
\begin{equation}\label{eq:infty_sol}
    R(x) \rightarrow \frac{_s R^{l m }_{\text{in}}}{r_+} \frac{e^{-i \bar{\omega} x}}{x} + \frac{_s R^{l m }_{\text{out}}}{r_+^{2s+1}} \frac{e^{i \bar{\omega} x}}{x^{2s+1}}\,,
\end{equation}
where $\bar{\omega} = \omega r_+$.  
To evaluate the transmission coefficient, we extract the coefficient $_s R^{l m}_{\text{in}}(\omega)$. The normalization of the scattering problem is set by choosing $a_0 = 1$ in \eqref{eq:frob}, which is equivalent to  
\begin{equation}  
    | _s R^{l m }_{\text{hole}}|^2 = (2 r_+ )^{2s} (a^2 - M^2)^2\,.
\end{equation}
This leads to the transmission coefficients for different spin fields
\begin{equation}\label{eq:GBF}
   \Gamma_{l m}^s = \delta_s | _s R^{l m}_{\text{in}}|^{-2}\,,
\end{equation}
where  
\begin{equation}
    \delta_s = - i e^{i \pi s} \bar{\omega}^{(2 s - 1)} \left(\frac{1}{2}\right)^{1-2s} \frac{\Gamma\left(1-s+i2\dfrac{\varpi}{\tau}\right)}{\Gamma\left(s+i2\dfrac{\varpi}{\tau}\right)} \tau\,.
\end{equation}
The near-horizon solution \eqref{eq:frob} is used as a boundary condition for numerically integrating the radial Teukolsky equation \eqref{eq:Teuk2} up to large distances where the asymptotic form is given bay \eqref{eq:infty_sol}. The numerical procedure starts at a radius $r_+ + \varepsilon$, or equivalently $x = \varepsilon / r_+$, where $\varepsilon \to 0$. Typically, we take $\varepsilon = \mathcal{O}(10^{-5})$. The solution is then evolved outwards to a sufficiently large radius, $r \gg r_+$.

\noindent At this asymptotic distance, the coefficients $_s R^{l m }_{\text{in}}$ and $_s R^{l m }_{\text{out}}$ are extracted by matching the numerical solution to the expected asymptotic form given in Eq.~\eqref{eq:infty_sol}. The GBF is subsequently determined using Eq.~\eqref{eq:GBF}.  
As we will show, this method (that for now on, in this paper, will be dubbed \emph{direct method}) exhibits significantly greater numerical stability compared to the Schr\"odinger-like approach (referred to as \emph{Chandrasekhar method}).\\
\noindent It has to be mentioned that other methods exist, as for example, the analytic one proposed in \cite{Bonelli:2021uvf}. Nonetheless, the study of those methods goes beyond the scope of this article and will be taken into account in future works.

\section{Results}\label{sec:results}

\begin{figure}[t]
    \centering
    \includegraphics[width=0.5\linewidth]{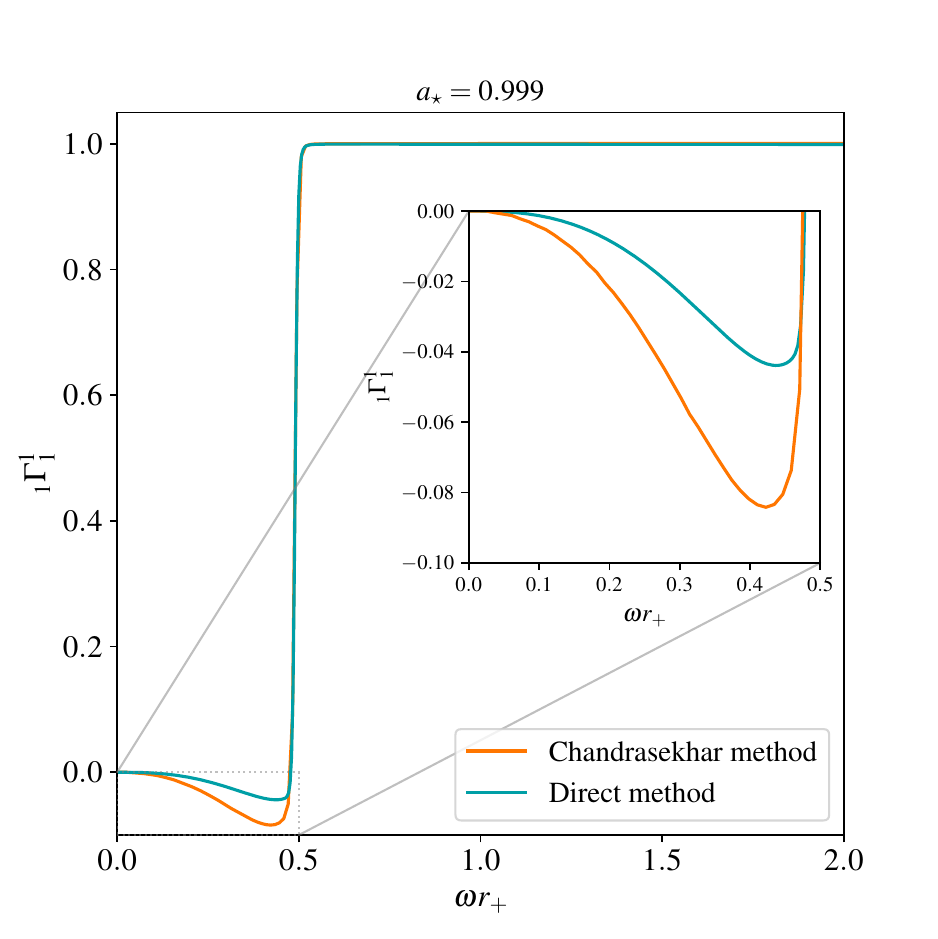}
    \caption{GBFs of a massless vector field mode $l=m=1$ for a Kerr BH with $a_\star=0.999$ as function of $\omega r_+$. In yellow the one computed with the \emph{Chandrasekhar method}, in blue the one computed from the Teukolsky equation with the \emph{direct method}.}
    \label{fig:fig2}
\end{figure}

\noindent Despite the appeal of the method described in Sec.~\ref{sec:sch_form}, which allows the application of various techniques developed for solving Schr\"odinger-like equations, its numerical implementation presents challenges. Specifically, the presence of singularities in the effective potential as $\epsilon \to 0$ introduces numerical errors.  
To rigorously compare the numerical accuracy of these methods, we independently computed the GBFs using both approaches described in Secs.~\ref{sec:sch_form} and~\ref{sec:bet_form}. In the first case, we followed the prescriptions outlined in Refs.~\cite{Arbey:2019mbc,Arbey:2021mbl} and made use of the Mathematica notebook based on the \emph{Chandrasekhar method} provided in the $\tt BlackHawk$ folder and:$~scripts/greybody\_scripts/greybody\_factors/spin\_1.m$. 
As oulined in Sec.~\ref{sec:bet_form}, we numerically computed $\,_s\Gamma^m_{l}$ for different field spins, including modes up to $l=4$. 
This was achieved by solving Eq.~\eqref{eq:Teuk2} with $\,_sQ^{m}_l$ and expanding Eq.~\eqref{eq:Near} up to fifth order, considering various black hole spin values. 
We sampled $600$ equally spaced points between $\omega r_+ = 0$ and $\omega r_+ = 2.4$.  
This upper frequency limit is sufficiently large for all relevant transmission coefficients to approach unity, ensuring accuracy in our calculations. Furthermore, higher-frequency modes are Boltzmann-suppressed in the Hawking emission spectra, making their contributions negligible.

\noindent An example of the discrepancies between the two methods is reported in Fig.~\ref{fig:fig2} where it is shown the GBF  of a photon field ($s=1$) in leading emitted mode $l=m=1$ as emitted by a nearly extremal (a=0.999) Kerr BH. 

\noindent As shown in Fig.~\ref{fig:fig2}, at low energies, the method described in Sec.~\ref{sec:sch_form} (orange line) overestimates the well-established results for superradiant amplification, which has a value of $\sim 4\%$, as reported in Ref.~\cite{Rosa:2016bli,Teukolsky:1974yv,Brito:2015oca} and references therein. 
In contrast, the direct solution method (green line) correctly reproduces these expected values.  
In general, the presence of $\epsilon \neq 0$ leads to an overestimation of the absolute value of the GBFs at low energies. 
We numerically verified that fixing $\epsilon$ (as done in Refs.~\cite{Arbey:2019mbc,Arbey:2021mbl}) results in increasing GBF discrepancies as the spin parameter $a_\star$ increases. 
Additionally, we explored the impact of reducing $\epsilon$ and found that the discrepancies worsened while the convergence of the solutions became increasingly unstable. 

\noindent However, these deviations remain small compared to the rapid variations in the GBF and are confined to the superradiant regime.  
Despite their localized nature, these deviations can influence predictions and constraints in scenarios where the low-energy spectrum dominates, potentially altering conclusions drawn in such cases.
To assess the impact of the overestimation of gray-body factors in the Chandrasekhar method, we examine the Hawking emission rate given in Eq.~\eqref{eq:PHR}.  
For a Kerr black hole, the parameter $k$ in Eq.~\eqref{eq:PHR} is given by $k = \omega - m\Omega_H$, which becomes negative in the superradiant regime ($\omega < m\Omega_H$). Simultaneously, the GBFs also become negative in this regime, ensuring that the overall emission rate remains positive.  
However, the denominator in Eq.~\eqref{eq:PHR} naturally favors low-frequency modes, which can lead to an unphysical enhancement of the spectrum for large values of $a_\star$ when GBFs are computed using the method in Sec.~\ref{sec:sch_form}.
\begin{figure}[t]
    \centering
    \includegraphics[width=\linewidth]{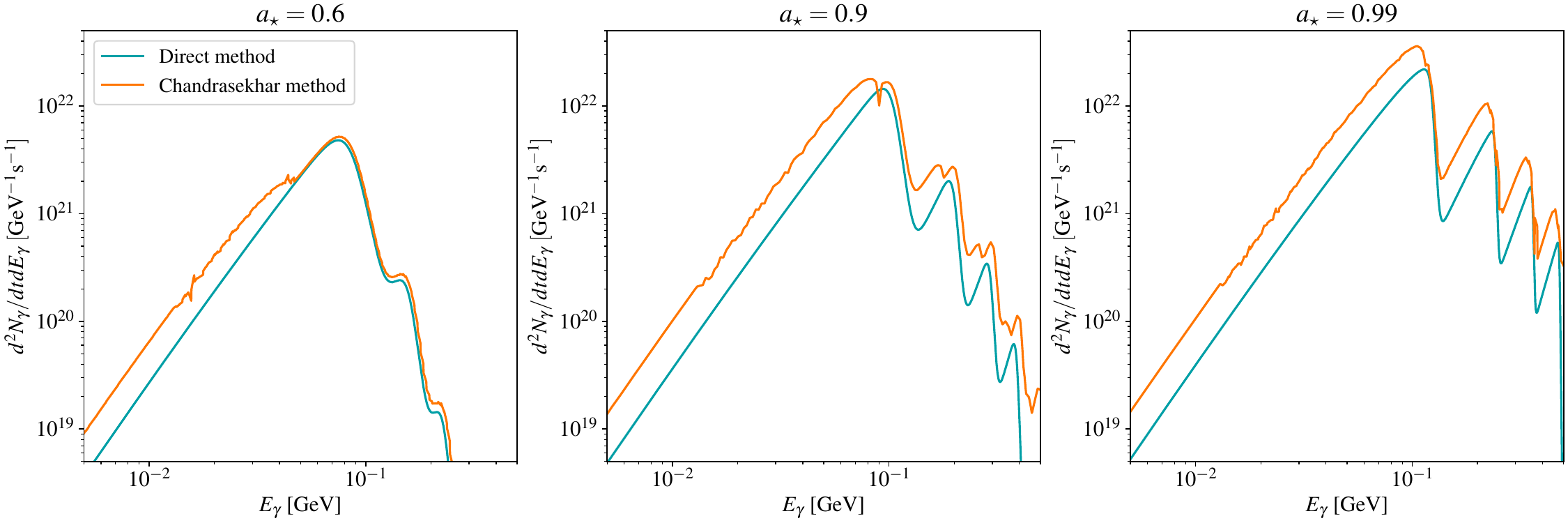}
    \caption{Primary emitted photon emission for a Kerr BH of mass $10^{15}$ g computed with Schr\"odinger-like equation with \emph{Chandrasekhar method} (in orange) and with the GBFs computed from the Teukolsky equation with the \emph{direct method} (in green). In the above panel from left to right $a_\star=0.6, 0.9, 0.99$.}
    \label{fig:fig3}
\end{figure}
Fig.~\ref{fig:fig3} presents the primary photon emission spectra for a Kerr black hole with a mass of $10^{15}$ g to illustrate this unphysical enhancement. The spectra are computed using the Chandrasekhar method (orange) and the direct method (green) for different values of the black hole spin parameter $a_\star$.  
As expected, the discrepancy increases with larger $a_\star$, reaching a factor of approximately two for $a_\star = 0.99$. Furthermore, we observe that the Chandrasekhar method exhibits significant numerical instability, particularly for $a_\star = 0.9$, where spurious peaks appear that are absent in the correct Hawking spectrum.  
Finally, we emphasize that this behavior is consistent across all field spins.

\begin{figure}[t]
    \centering
    \includegraphics[width=0.5\linewidth]{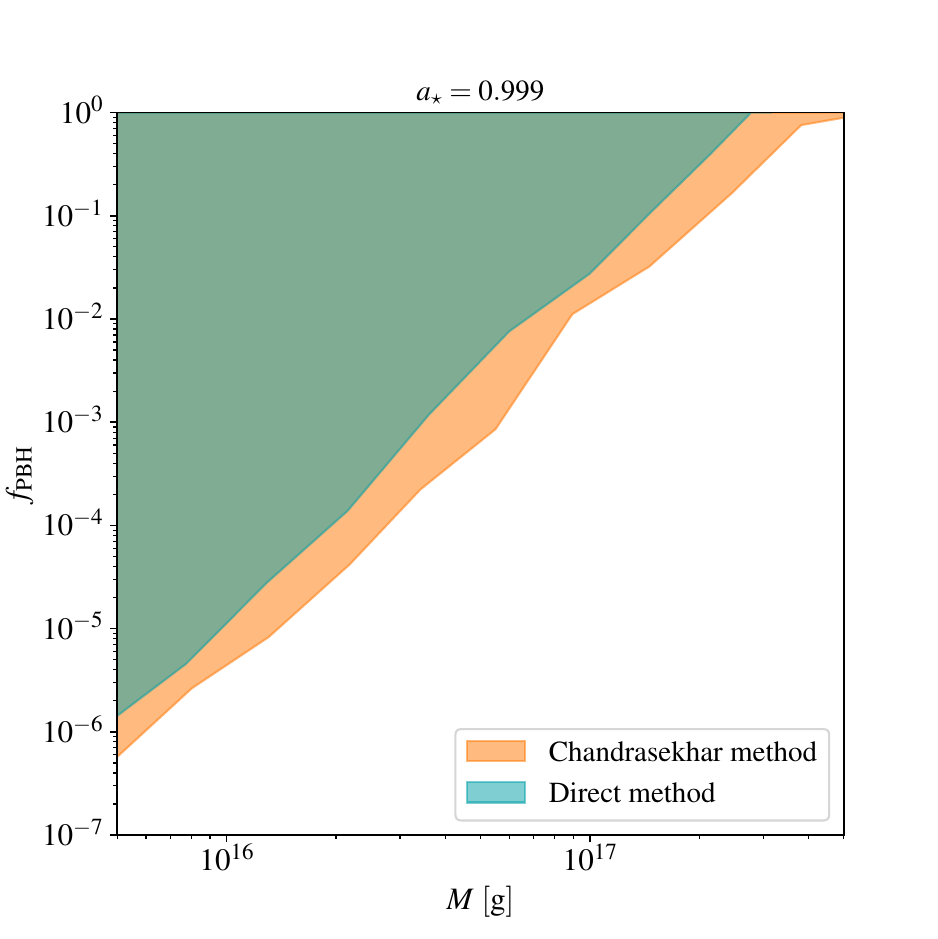}
    \caption{Constraints on the dark matter fraction in form of Primordial Black Holes $f_{\rm PBH}$ as function of the mass coming from the non-observation of a gamma-ray flux, assuming that the black hole population has a close-to-maximally angular momentum, with $a_\star=0.999$. The green region is the limit obtained with the \emph{direct method} computation of the gray-body factors, while the orange one is the bound using the factors obtained from the Schr\"odinger-like \emph{Chandrasekhar method}.}
    \label{fig:fig4}
\end{figure}

\noindent Since this overestimation is present in the latest versions of \texttt{BlackHawk}~\cite{Arbey:2019mbc,Arbey:2021jif}, it is important to assess its impact on phenomenological studies that have considered Kerr PBHs and relied on \texttt{BlackHawk}.  
As an example of the consequences of this overestimation in the Hawking spectrum for photons, we compute the constraints on the fraction of primordial black holes (PBHs) that form the dark matter $f_{\rm PBH}$ derived from the diffuse $\gamma$-ray background using both methods in the case of nearly extremal PBHs. 
Our results are presented in Fig.~\ref{fig:fig4}. We present the excluded regions computed using the correct and inaccurate GBFs for photons.
We find that for $a_\star = 0.999$, the constraint is weakened by a factor of approximately two compared to the results in Ref.~\cite{Arbey:2020yzj,Arbey:2019vqx}.  

\noindent Furthermore, these discrepancies extend beyond static properties of the black hole, such as instantaneous spectra and their associated constraints. They also affect the dynamical evolution of a spinning BH. Since GBFs play a crucial role in defining the depletion functions $f$ and $g$, which govern the time evolution of a black hole~\cite{Page:1976df,Page:1976ki,Page:1977um,Chambers:1997ai,Taylor:1998dk}, their accuracy is essential. While a detailed investigation of these effects is beyond the scope of this brief work, we intend to address them in a future study.  
Moreover, we are working on an update of \texttt{BlackHawk} which will incorporate the corrected GBFs \cite{BlackHawk_v3}.

\section{Conclusions}\label{sec:conc}

\noindent The methodologies for calculating GBFs in both non-rotating and rotating black hole spacetimes are of paramount importance in understanding the Hawking emission spectra, which in turn inform our knowledge of black hole thermodynamics and their astrophysical implications. The Schr\"odinger-like formalism, widely used for its simplicity and computational efficiency in spherically symmetric cases, has proven effective in scenarios involving static or Schwarzschild black holes. However, this approach encounters significant challenges in the context of Kerr black holes, particularly in the superradiant regime characterized by high angular momentum.

\noindent The reliance on Chandrasekhar tortoise coordinates, while enabling the Schr\"odinger-like representation, introduces numerical errors that can manifest as inaccuracies in the computed GBFs. These errors become increasingly pronounced for black holes with large spin parameters, where the superradiant amplification of certain modes is sensitive to the details of the numerical implementation. Such inaccuracies not only affect the GBFs but also propagate into the predictions of the Hawking emission spectra, potentially leading to deviations in the expected observational signatures.

\noindent To overcome these limitations, alternative methods that avoid the Chandrasekhar tortoise coordinate have been developed and analyzed. These approaches, leveraging a re-scaled radial coordinate and Frobenius-like expansions, offer a robust framework for addressing the complexities of rotating black hole spacetimes. Notably, they provide improved accuracy in computing GBFs for Kerr black holes, particularly in regimes where superradiant effects dominate and low-energy phenomena are crucial. This refinement is vital for ensuring reliable predictions in studies of black hole phenomenology, including those exploring the role of primordial black holes as dark matter candidates or as sources of detectable high-energy emissions.

\noindent The results presented highlight the importance of methodological precision and underscore the need for further optimization of computational techniques in this domain. Future research should aim to validate these approaches across a broader range of black hole parameters and field spins, ensuring their applicability to diverse astrophysical and cosmological scenarios. Moreover, integrating these improved methods into widely-used computational tools will enhance their accessibility and facilitate more accurate analyses in both theoretical and observational contexts. Such advancements are essential for advancing our understanding of black holes and their role in the universe.

\section*{Acknowledgements}

\noindent YFPG acknowledges financial support by the Consolidaci\'on Investigadora grant CNS2023-144536 from the Spanish Ministerio de Ciencia e Innovaci\'on (MCIN) and by the Spanish Research Agency (Agencia Estatal de Investigaci\'on) through the grant IFT Centro de Excelencia Severo Ochoa No CEX2020-001007-S. \\
\noindent MC acknowledges support from the Istituto Nazionale di Fisica Nucleare (INFN) through the Commissione Scientifica Nazionale 4 (CSN4) Iniziativa Specifica ``Quantum Fields in Gravity, Cosmology and Black Holes'' (FLAG). M.C. acknowledges support from the University of Trento and the Provincia Autonoma di Trento (PAT, Autonomous Province of Trento) through the UniTrento Internal Call for Research 2023 grant ``Searching for Dark Energy off the beaten track'' (DARKTRACK, grant agreement no.\ E63C22000500003).

\bibliographystyle{apsrev4-1}
\bibliography{main.bib}

\end{document}